# Implementation of the direct evaluation of strains in a frequency-based image analysis code for random patterns

J. Molimard *[*, a]

[a] Claude Goux Laboratory , UMR CNRS 5146, University of Lyon, Health Engineering Center, Ecole des Mines de Saint-Etienne, 158 cours Fauriel, 42023 Saint-Etienne, France

**ABSTRACT**

*A new approach for decoding displacements from surfaces encoded with random patterns has been developed and validated. The procedure is based on phase analysis of little zones of interest. Resolution in standard conditions (32×32 pixels²) is 2/100th pixel, for a spatial resolution of 9 pixels. Here we adapt new concepts proposed by Badulescu (2009) on the grid method to random patterns for the direct measurement of strains. First metrological results are encouraging: resolution is proportional to strain level, being 9% of the nominal value, for a spatial resolution of 9 pixels (ZOI 64×64 pixels²). Random noise have to be carefully controlled. A numerical example shows the relevance of the approach. Then, first application on a carbon fiber reinforced composite is developed. Fabric intertwining is studied using a tensile test. Over-strain are clearly visible, and results connect well with previous studies.*

**Highlight:**

---

\* molimard@emse.fr, phone 33 4 77426648, fax 33 4 77420249, www.emse.fr/~molimard

- sub-pixel interpolation is achieved using phase component
- strain maps can be extracted without derivation of a displacement map
- small speckles and gray level noise averaging give best results

## 1. INTRODUCTION

Digital image correlation is one of the most diffused image processing technique among experimental mechanics community [1]. The system has been extended to different cases and firstly on warp surfaces (stereocorrelation) [2, 3]. But basically, one of the most interesting problems in DIC is the sub-pixel detection. Usually, authors use either correlation peak interpolation either sub-pixel deformation of the target sub-image. The deformation hypothesis can be a rigid body motion or a complete transformation function of the zone of interest (ZOI) [4]. Recently, Hild [5] proposed an original work, based on a correlation algorithm coupled with Finite Element based transformation functions, allowing for strongly regularized displacement fields with solid mechanics assumptions. In all these cases, the assumptions on the interpolation level are believed to have a direct influence on metrological properties of the correlation core [6].

Beside the classical DIC approach, the grid method, even if less used, has rather comparable features: it is basically an image processing technique allowing for in-plane displacement fields. In the previous case, images are random patterns; in the latter, surface signature is a periodic grid. In this case, displacements derive from a spatial phase extraction [7]. The choice of surface encoding should orient the user toward one method or the other, but in fact, the nature of the pattern is not crucial: DIC has been successfully applied to periodic pattern for a very long time [8], and a random pattern can be seen has the superimposition of many frequencies. This last remark have been made recently by some authors [9, 10, 11] and will be developed hereafter. Random pattern are of more practical use than periodic pattern for two basic reasons: first, periodic pattern generation and transfer is not as easy as someone could think. Second, it is almost impossible to generate a periodic pattern on a non-flat surface – and to develop a 3D surface grid method, even if industrial demand for measurement on real structures is high.

Finally, one should note that DIC is a genuine large-strain approach because it is based on re-correlation of decorrelated informations whereas Grid technique is a genuine small perturbation method because the phase difference is merging initial and final state of the investigated object, and because phase information is fairly more sensitive than amplitude. Consequently, it could be of great interest to adapt advances made in the context of grid techniques to random patterns, and to compare the results to those obtained by classical image correlation technique. We noted here the following specific "phase culture" items: camera distortion can be evaluated with a single grid image [12], shape and shape variation can be detected in very good conditions using a video-projector and a camera [13]. Last, derivation can be derived analytically and the grid technique becomes sensitive to strain rather than displacements [14]. This item should bring to such method a breakthrough compared to DIC techniques; it will be developed in the following.

## 2. 2D DISPLACEMENT METHOD USING FREQUENCY-BASED ANALYSIS

### 2.1 Digital Image Correlation Principle

Assuming a reference image im0 described by $f(x,y)$, a deformed image im1 of im0 after a small strain is described by $g(x,y)$ by the following equation:

$$g(x, y) = f(x - \delta_x, y - \delta_y) + b(x, y) \quad (1)$$

where $\delta x$ and $\delta y$ are the components of the displacement of im1 and $b(x,y)$ the noise measurement. A way to find $\delta x$ and $\delta y$ is to maximize the function $h$ defined by:

$$h(x, y) = (g * f)(x, y) = \int_{-\infty}^{+\infty} \int_{-\infty}^{+\infty} g(\xi, \eta) f(\xi - r, \eta - s) d\xi d\eta \quad (2)$$

where * denotes the cross-correlation product. The obtained $r$ and $s$ correspond to the maximal probability of displacement ($\delta x$, $\delta y$). This method can be applied in Fourier space using Fast Fourier Transform function, noted FFT2D. Equation 2 becomes:

$$g * f = FFT2D^{-1}(FFT2D(g) \overline{FFT2D(\tilde{f})}) \quad (3)$$

where the overline denotes the complex conjugate, and $\tilde{f}(x, y) = f(L_x - x, L_y - y)$ where $L_x$ and $L_y$ are the dimension of the domain. Equation 2 and 3 are used at a local scale, on typical 32×32 zone of interest (ZOI). The work is repeated all over the map, giving a final displacement chart.

Now, classical digital image correlation performs the cross-product either in Fourier or real space. More refined approachs exist by the way: for example, the signal could be normalized respect to the mean local intensity, and/or the mean local contrast [6]. The cross-correlation peak is commonly interpolated in order to reach a sub-pixel displacement accuracy. The interpolation function is not the same for all the implementations. It could be for example a Gaussian or polynomial function. Basically, this choice doesn't have a strong theoretical basis, and authors have an empirical approach. We will propose hereafter an alternative to this weak point in the image correlation approach.

**2.2 Sub-Pixel algorithm**

The sub-pixel algorithm is based on the phase estimation of each ZOI. Because Fourier Transform requires continuity, in particular at the boundaries, the ZOI is windowed using a bi-triangular function. So far, the algorithm is an extension of the windowed Fourier Transform (WFT) algorithm proposed by Surrel [15]. As shown Fig. 1, in the frequency domain, each couple of frequencies is characterized by an amplitude and a phase. This phase is proportional to the displacement normal to the corresponding fringe direction. Note that the phase is defined only if the amplitude is higher than zero. Then, in absence of any phase jump, displacements can be related to any defined phase using the relationship:

$$\left\{ \begin{array}{c} \vdots \\ \Delta \phi_\theta^i \\ \vdots \end{array} \right\} = \underbrace{\begin{bmatrix} \vdots & \vdots \\ \dfrac{2\pi}{p_\theta^i}\cos\theta^i & \dfrac{2\pi}{p_\theta^i}\sin\theta^i \\ \vdots & \vdots \end{bmatrix}}_{A} \left\{ \begin{array}{c} \delta_x \\ \delta_y \end{array} \right\} \qquad (4)$$

Displacements can then be derived from eq. 4 using the pseudo-inverse of A. This operation is possible if $\det((A^t A)) \neq 0$. In practice, this means that at least two phases along two different directions exist.

$$\begin{Bmatrix} \delta_x \\ \delta_y \end{Bmatrix} = (A^t A)^{-1} A^t \begin{Bmatrix} \vdots \\ \Delta \phi_\theta^i \\ \vdots \end{Bmatrix} \quad (5)$$

Practical problem of this approach is that the signal to noise ratio is weak for each couple of frequencies. Then, the quality of the measurement is obtained by averaging all the available information throw the pseudo-inverse function.

One should note also that a phase jump can occur in the Fourier domain. Even if a specific treatment should be developed, it sounds better to first use a pixel correlation algorithm. This ensures that the two ZOIs will be as superimposed as possible and that the fringe order is zero for any point and any couple of frequencies. No deformation of the ZOI is proposed here, considering that target applications will be in the small transformation domain.

## 2.3 Characterization

The method has been characterized using a simulation of a rigid body translation. A single ZOI is generated and translated from -0.5 to 0.5 pixel, in presence of noise or not. The ZOI has been over-sampled 10 times. 10 translation cases equally spaced are studied, and for each displacements 60 pairs of ZOI are sampled. Typical results are shown in fig. 2 for a 32×32 region of interest, 12bit camera, with a 31 gray level white noise or not. One should note first that the error has the same amplitude as with image correlation techniques (Table 1), but the bias is fairly lower in this case. This should be one major advantage of this method.

Spatial resolution of the method is estimated using the following procedure: two independent noise distributions are added to the same speckle image. Then, displacements between the two situations are calculated, giving a displacement error map. The autocorrelation function of this error map is only affected by the displacement extraction procedure, so its size characterize the spatial resolution of the displacement extraction procedure. For the 32×32 window, effective spatial resolution is a diameter of 9 pixels (at 50% attenuation). This result should be surprising, but it can be partially explained by the use of the bi-triangular window: in 1D (the classical WFT algorithm) spatial resolution is half the window size. Then, it is worth noting that these values are very interesting compared with others zero-order image correlation algorithms.

## 3. DIRECT DERIVATION IMPLEMENTATION

### 3.1 Basic principle

Recently, Badulescu and al. [14] proposed a new derivation procedure for grid techniques. The technique is twofold: first, it consists in deriving the wavelet used for the phase extraction. Authors claim that this approach leads to better results than deriving the displacement map using a classical low-pass filter coupled with a least-square derivation (called hereafter the classical procedure). In the phase extraction procedure, the phase is defined as:

$$\phi_\theta(x,y) = arg(R_\theta + iJ_\theta) = arg\left(\int_{-\infty}^{+\infty}\int_{-\infty}^{+\infty} s(\xi,\eta) \times g(\xi-x, \eta-y) \times \exp\left(-i\frac{2\pi}{P_\theta}(\xi\cos\theta + \eta\sin\theta)\right) d\xi\, d\eta\right) \qquad (6)$$

where $s(\zeta, \eta)$ is the signal and $g(\zeta-x, \eta-y)$ is a windowing Gaussian function. Note that this function is different than the one used previously for displacement measurement. If considering the derivation of $U_\theta$ along $\vec{x}$,

$$\frac{\partial \phi_\theta}{\partial x} = \left( \frac{\frac{\partial J_\theta}{\partial x} \times R_\theta - \frac{\partial R_\theta}{\partial x} \times J_\theta}{J_\theta^2 + R_\theta^2} \right) \tag{7}$$

$$\frac{\partial U_\theta}{\partial x} = -\frac{2\pi}{P_\theta} \Delta \left( \frac{\partial \phi_\theta}{\partial x} \right) \tag{8}$$

Last, the derivation of $J_\theta$ or $R_\theta$ along $x_\theta$ consists in deriving $g(\zeta, \eta)$ i.e.:

$$\frac{\partial J_\theta}{\partial x} = \int_{-\infty}^{+\infty} \int_{-\infty}^{+\infty} \frac{(x-\xi)}{2\pi \sigma^4} \times s(\xi, \eta) \times g(\xi-x, \eta-y) \times \exp\left(-i\frac{2\pi}{P_\theta}(\xi \cos\theta + \eta \sin\theta)\right) d\xi d\eta \tag{9}$$

Second, reference and deformed phase gradient maps are numerically superimposed to avoid noise propagation due to local variations in the grid signature. In each situation, authors show that the results are better if noise intensity is reduced using temporal averaging. Depending on the conditions, the resolution can be one order of magnitude lower than the classical procedure for a given spatial resolution.

Now, the grid technique can be seen as a particular case of the frequency-based image analysis of random patterns. So far, it is possible to generalize this new strain measurement procedure. The proposed implementation is based on:

1. the superimposition of the original and deformed images, using a linear interpolation of gray levels. Note that it is impossible here to deform directly the phase gradient maps.

Deformation of the intensity map is a major difference between the two approaches. In a first step, intensity maps are submitted to a simple translation. Then, a first evaluation of the strain is used to deform the intensity map, and procedure stop when convergence is achieved.

2. the expression of phase derivative according to (7),

3. instead of using direct expressions, a calibration procedure for 4 reference numerical deformations of the original ZOI is made (two translations, a rotation, and in-plane shear), giving an in-situ sensitivity. It is believed that this procedure is less sensitive to ZOI signature.

**3.2 Characterization**

The method has been characterized using a simulated deformation of a single ZOI. The ZOI is submitted to uniaxial strain at different levels ranging from $2.10^{-5}$ to $2.10^{-3}$. In order to characterize average behaviour, , 60 ZOI are generated for each strain level. The rigid body displacement is set to zero in order to avoid possible coupling effects. For each case, all the displacement gradients are recorded and mechanical strain tensor is reconstructed ($\varepsilon_{xx}$, $\varepsilon_{yy}$, $\varepsilon_{xy}$, $\omega$). The different cases are:

- ZOI varies from 32×32 and 64×64 $px^2$;
- noise on intensity from 0 to 31 gray levels;
- speckle radius from 2 to 16,
- encoding deepness from 8 to 16 bits.

Because of the high number of parameters, we use here a linear experimental test plane. Results are presented Table 2. Note that every parameter has a good enough confidence interval, and is possible to draw the main behavior. It was necessary to add 1st order interaction to have a good enough model. Finally, uncertainty on strain increases with the speckle size and with intensity noise, and decreases when ROI size and encoding increase. Note that speckle size and noise are strongly correlated, with a positive effect. Finally, results can be strongly improved by minimizing gray level noise using image averaging. Last, it is proved here that the use of small speckle is necessary to achieve good accuracy. Best results (64×64 px² ZOI, 2 px speckle radius, 0 GL noise, 16 bits encoding) gave a mean final uncertainty of 58 µm/m. In this case, the uncertainty is entirely random, with no bias. Dispersion is related to quantization error and to random process generation of the speckle pattern. Results are presented figure 4. The uncertainty increases with the load case, and finally, the relative uncertainty remains roughly constant (9 %).

Last, spatial resolution must be measured as explained before. Here, the value obtained for the 64×64 zone of interest is 9 pixels. This very low value can be explained by the use of a Gaussian windowing with $\sigma$ = 10.7 pixels.

## 4. EXAMPLES

### 4.1 Simulated experiment

First test on the sub-pixel algorithm has been conducted on a fake displacement map provided by GDR 2519 [6]. It consists in a sine-wave displacement field, as shown figure 4 encoded on a 8-bits intensity map, without noise. Here, the displacement amplitude has been set to 0.02 pixels, and spatial wavelength to 128 pixels. Root mean square error found both in $x$ and $y$ directions is 0.003 pixels. The strain map (fig.5) are focused on a limited area. Its obvious anyway that the values show a very good trend. If comparing classical derivation with the same spatial resolution, this new technique gives better results in absence of noise: 1.7 10$^{-4}$ for the proposed technique vs. 2.6 10$^{-4}$ for the classical one.

**4.2 Strain field a single-ply carbon-fiber reinforced composites**

Carbon fiber reinforced composites are commonly used for structural parts. The reinforcement can be delivered as UD tapes; mechanical properties are at the maximum, but handling is difficult. On the opposite, fabrics can be handled easier, but the intertwining of tows decreases their loading capacity. This effect is due to the development of local stresses. Then, it is important to characterize a weaving with a security coefficient corresponding to the local amplification of global stress. Because the geometrical description of a fabric is difficult, experimental study through an optical full-field technique is a good solution. This work has already been done by Lee [16] using interferometry technique outlining strain concentrations. Here, the idea is to propose a simpler optical set-up using a single camera. Direct derivation is a good way to catch very localized phenomena.

The result of a feasibility test is given figure 6 and 7. The specimen is a composite made of a single fabric carbon ply. The fabric unit cell is 8×8 mm². Applied load is 200 N. Extraction of data is made here with a 64×64 ZOI. Displacement maps show variations in conjunction with the tow interlacing. Strain map obtained on figure 7 had to be filtered for cosmetic reasons using a first a salt-and-paper filter, then a Gaussian low-pass filter with σ = 9 pixels. Results are in good agreement with former works on the same structure [16, 17]. It is shown that interlacing of tows implies periodic over-strain with band (banana) shapes. The shear strain is maximum in the resin rich region at the crossing of longitudinal and transverse tows. Here, the final spatial resolution is 13 pixels. This value is very low compared to other geometric methods, such as digital image correlation, or simply to a derivation post-processing.

## 5. CONCLUSION AND PERSPECTIVES

Strain evaluation is a very critical task for mechanical engineers. Usual geometrical methods (image correlation, grid techniques) classically need a post-processing based on a low-pass filter combined with least square fit. Recently, we proposed a frequency-based image analysis code for random patterns. This code has been extended to direct strain evaluation using a phase derivation kernel.

First results show very promising performances if the noise level is controlled. Main advantage of the method should be its final spatial resolution. Two examples illustrate the first applications of the technique. The first one is a simulated experiment; the second one is on a carbon fiber reinforced composite. The fabric interlacing produces local strain peaks, visible with the described procedure.

## ACKNOWLEDGEMENTS

Author thanks Saint-Etienne Metrope for granting this work

**TABLE CAPTION**

Table 1. Error quantification for a 32×32 window.

Table 2. Error sensitivity to parameters

Table 1. Error quantification for a 32×32 window.

|  | 0 GL noise | 31 GL noise |
| --- | --- | --- |
| Resolution (in pixels) | 0.025 | 0.035 |
| Bias (in pixels) | 0.0044 | 0.0056 |

Table 2. Error sensitivity to parameters

| | | Value | | Confidence level | |
|---|---|---|---|---|---|
| Mean value | $\beta_0$ | $1.07\ 10^{-1}$ | | 98.4 % | ++++ |
| ROI size effect | $\beta_1$ | $-7.85\ 10^{-2}$ | ▲ | 98.3 % | ++++ |
| Speckle size effect | $\beta_2$ | $1.80\ 10^{-1}$ | ▼ | 98.0 % | ++++ |
| Encoding effect | $\beta_3$ | $-7.40\ 10^{-2}$ | ▲ | 98.5 % | ++++ |
| Intensity noise effect | $\beta_4$ | $1.1\ 10^{-1}$ | ▼ | 98.0 % | ++++ |
| ROI size / Speckle size interaction | $\beta_{12}$ | $7.81\ 10^{-4}$ | ▼ | 13.5 % | - |
| ROI size / Encoding interaction | $\beta_{13}$ | $-1.89\ 10^{-3}$ | ▲ | 73.1 % | +++ |
| ROI size / Intensity noise interaction | $\beta_{14}$ | $-7.97\ 10^{-2}$ | ▲ | 97.8 % | ++++ |
| Speckle size / Encoding interaction | $\beta_{23}$ | $-3.86\ 10^{-3}$ | ▲ | 93.6 % | ++++ |
| Speckle size / Intensity noise interaction | $\beta_{24}$ | $1.88\ 10^{-1}$ | ▼ | 97.8 % | ++++ |
| Encoding / Intensity noise interaction | $\beta_{34}$ | $-7.17\ 10^{-2}$ | ▲ | 98.3 % | ++++ |

**FIGURE CAPTIONS**

Figure 1. Basic principle of sub-pixel algorithm.

Figure 2. Quality of identified displacements for a pure translation without and with noise.

Figure 3. Relative error on strain for the optimal test case (Spatial resolution 9 pixels).

Figure 4. In-plane displacement maps for a simulated in-plane displacement (in pixels).

Figure 5. In-plane strain maps for a simulated case. a/proposed procedure b/classical approach (in m/m).

Figure 6. Displacement maps obtained on a 1-fabric ply composite coupon under tension (in pixels).

Figure 7. Strain maps obtained on a 1-fabric ply composite coupon under tension a/$\varepsilon xx$, b/$\varepsilon xy$. Full scale is 10-3 (in m/m).

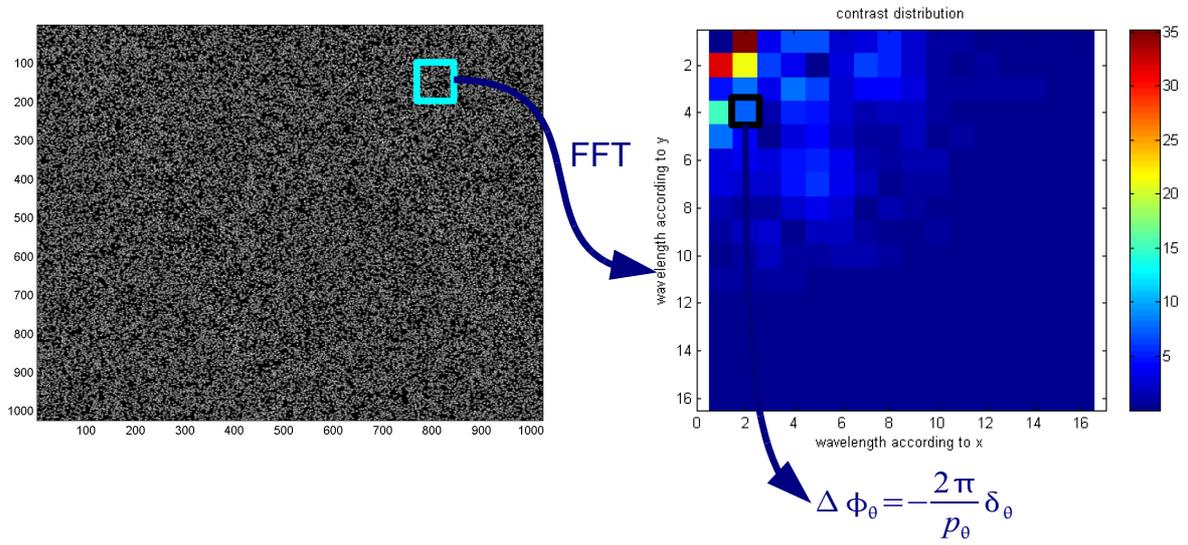

Figure 1. Basic principle of sub-pixel algorithm.

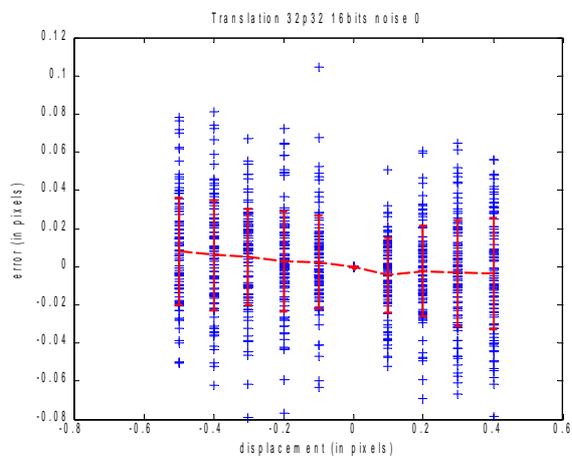 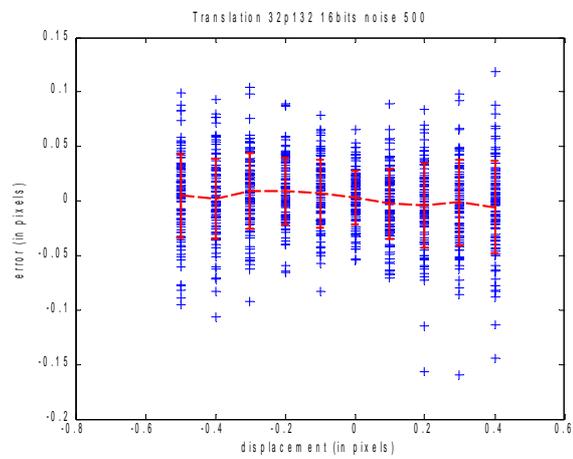

Figure 2. Quality of identified displacements for a pure translation without and with noise.

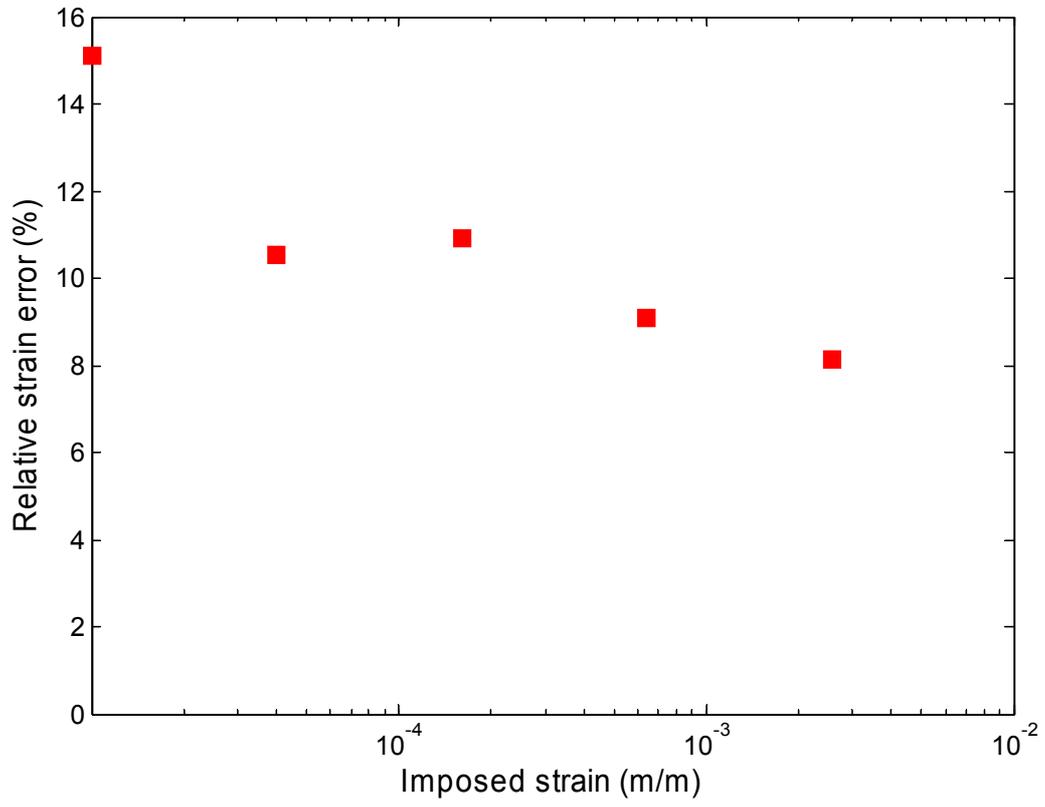

Figure 3. Relative error on strain for the optimal test case (Spatial resolution 9 pixels).

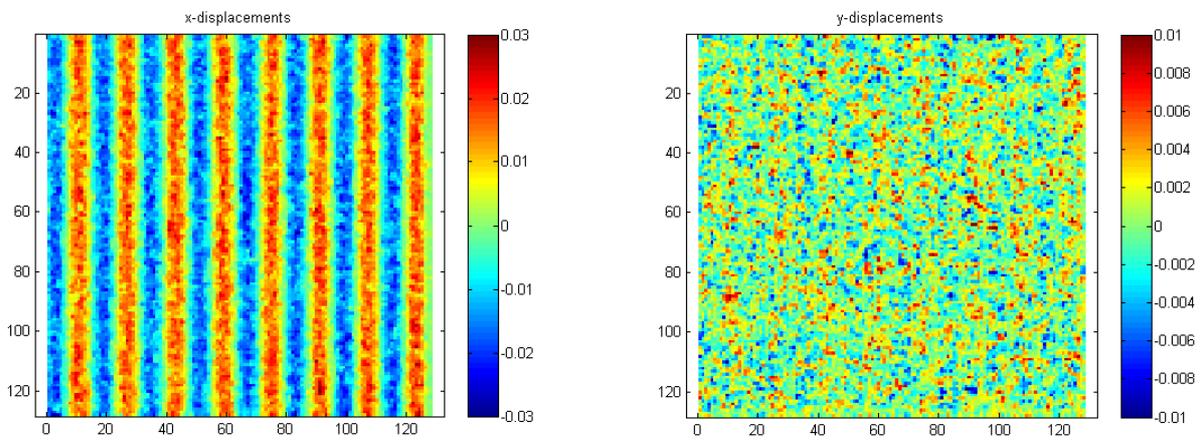

Figure 4. In-plane displacement maps for a simulated in-plane displacement (in pixels).

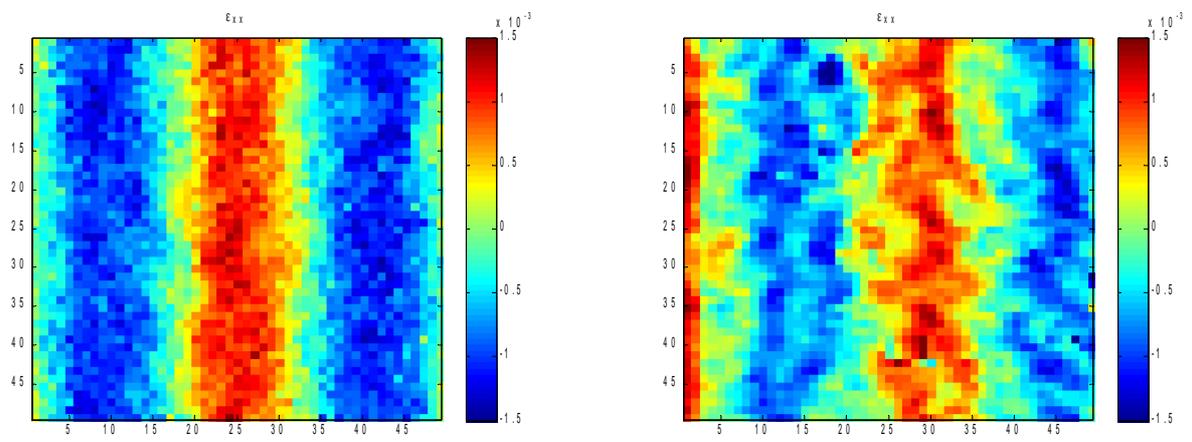

Figure 5. In-plane strain maps for a simulated case. a/proposed procedure b/classical approach (in m/m).

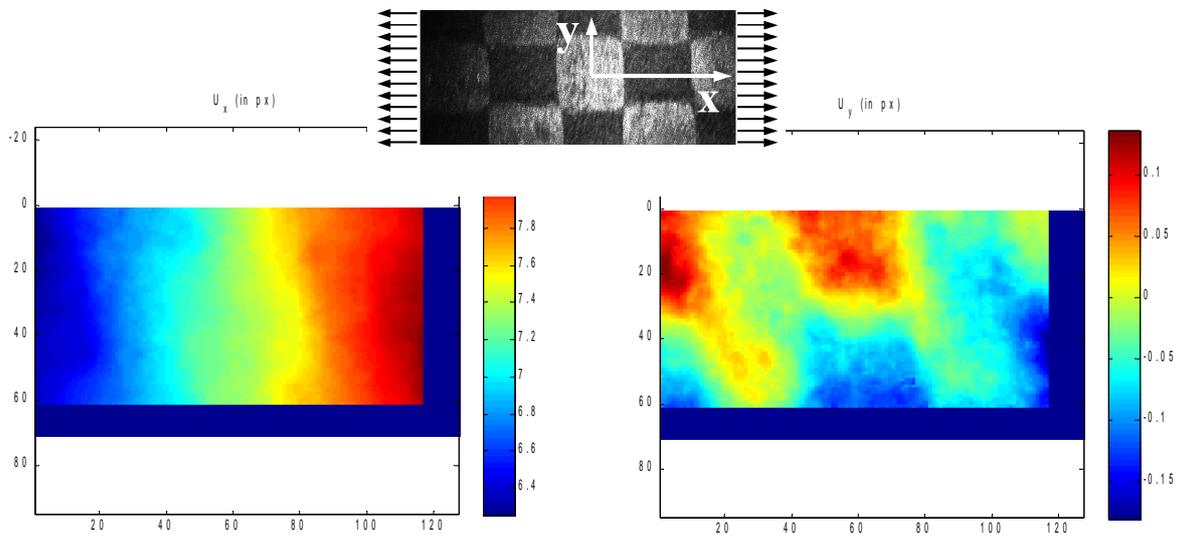

Figure 6. Displacement maps obtained on a 1-fabric ply composite coupon under tension (in pixels).

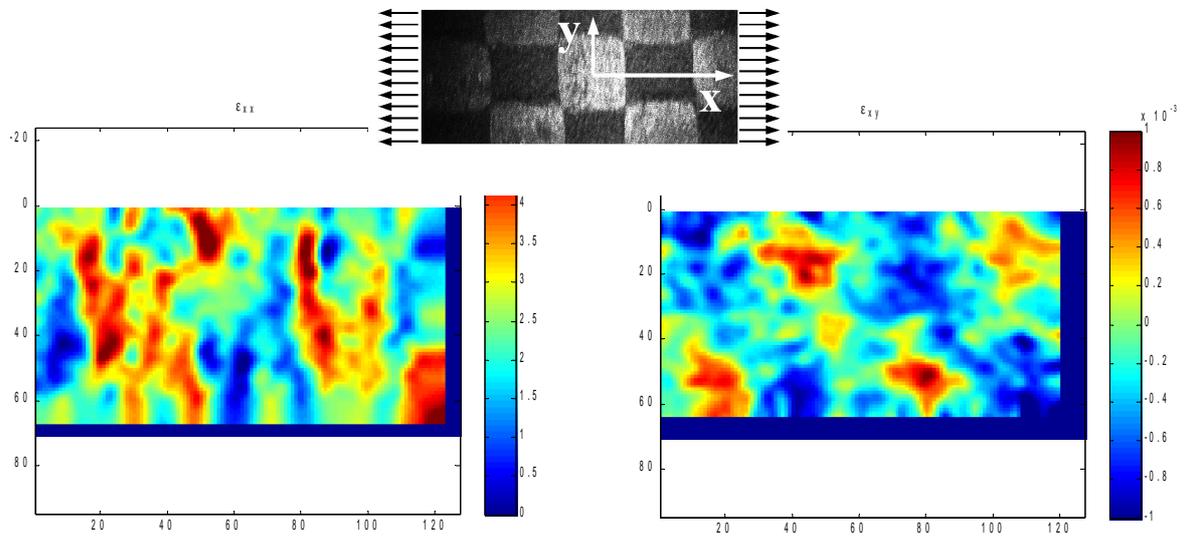

Figure 7. Strain maps obtained on a 1-fabric ply composite coupon under tension a/$\varepsilon_{xx}$, b/$\varepsilon_{xy}$. Full scale is $10^{-3}$ (in m/m).